\documentclass[nofootinbib,twocolumn,preprintnumbers]{revtex4-1}
\usepackage[dvips]{graphicx}
\usepackage{amsmath,amsthm,amssymb,multirow}
\begin{document}

\def\lsim{\mathrel{\rlap{\lower4pt\hbox{\hskip1pt$\sim$}}
    \raise1pt\hbox{$<$}}}
\def\gsim{\mathrel{\rlap{\lower4pt\hbox{\hskip1pt$\sim$}}
    \raise1pt\hbox{$>$}}}
\newcommand{\vev}[1]{ \left\langle {#1} \right\rangle }
\newcommand{\bra}[1]{ \langle {#1} | }
\newcommand{\ket}[1]{ | {#1} \rangle }
\newcommand{\ev}{ {\rm eV} }
\newcommand{\kev}{{\rm keV}}
\newcommand{\mev}{{\rm MeV}}
\newcommand{\gev}{{\rm GeV}}
\newcommand{\tev}{{\rm TeV}}
\newcommand{\mpl}{$M_{Pl}$}
\newcommand{\mw}{$M_{W}$}
\newcommand{\Ft}{F_{T}}
\newcommand{\Zparity}{\mathbb{Z}_2}
\newcommand{\BLambda}{\boldsymbol{\lambda}}
\newcommand{\be}{\begin{eqnarray}}
\newcommand{\ee}{\end{eqnarray}}
\newcommand{\met}{\;\not\!\!\!{E}_T}

\title{Searching for Direct Stop Production in Hadronic Top Data at the LHC}
\author{David E. Kaplan$^1$, Keith Rehermann$^2$, Daniel Stolarski$^{1,3}$}
\affiliation{$^1$Department of Physics and Astronomy, Johns Hopkins University, Baltimore, MD 21218\\ 
$^2$Massachusetts Institute of Technology, Cambridge, MA 02139\\ 
$^3$Center for Fundamental Physics, Department of Physics, University of Maryland, College Park, MD 20742}
\begin{abstract}
We argue that evidence can be uncovered for stops between 300-600 GeV in 5 fb$^{-1}$ of 7 TeV proton-proton collisions if they decay into top quarks and light neutral particles.  We also show that with 20 fb$^{-1}$ of 8 TeV running, discovery or exclusion can be made in large regions of parameter space. Events with a fully hadronic top/anti-top pair and a pair of invisible decay products can be identified with the top-tagging of a fat jet, a single $b$-tag, a missing transverse energy cut, as well as other kinematic cuts to reduce backgrounds with real top quarks in them.  Such cuts obliterate the background suggesting discovery can be made with a handful of events.
\end{abstract}

\preprint{MIT-CTP 4374, UMD-PP-012-007}

\maketitle

\section{Introduction}

One of the central questions in particle physics since the 1970s has been whether or not he electroweak scale is ``natural''~\cite{Susskind:1978ms, 'tHooft:1980xb}.  A theory that is not fine-tuned should have physics beyond the standard model that cuts off quadratically divergent loop contributions to the Higgs mass, and lives at roughly the same scale.  Estimating using a naive energy-momentum cut off, the top one-loop contribution is the largest, and thus especially requires new light states (or other physics) to cancel its effect.  In supersymmetric theories, it is the superpartner of the top quark, the scalar stop, that cancels the top contribution to the electroweak scale at one loop.  Thus, if supersymmetry is responsible for naturalness, the stop should be accessible at the LHC, and there has been much theoretical interest in determining the near-term prospects for stop discovery~\cite{Plehn:2012ab,Han:2012:ab,Alves:2012cd}. 

Here we investigate methods of discovering stops which are pair-produced and then promptly decay to tops and long-lived neutral particles.  It is natural for the stop to be the lightest squark if the running masses of all of the squarks are degenerate at a high scale.  This is because the large top Yukawa coupling causes the stop masses to run down at lower energies.  In addition, large mixing between the two complex scalar stops (again due to the large Yukawa coupling) suppresses one of the masses.  In the same vein, the neutralino is naturally the lightest fermion - either the bino (superpartner of the hypercharge gauge boson), or perhaps the singlino.\footnote{The neutral particle could also be a gravitino.}  The simplest way to explain a Higgs mass between 115-125 GeV in supersymmetry is to extend the minimal model to include a gauge-singlet chiral superfield, the singlino being the fermionic component.  Thus, in this article, we explore the LHC's ability to discover a stop (`right-handed,' for simplicity) which decays into a neutralino, while all other superpartners are decoupled. 

While the LHC has not published any direct stop searches, the final state of tops with $\met$ has been searched for with 1 fb$^{-1}$ of data~\cite{Aad:2011wc}. This search places limits on the existence of fermionic top partners, but it is insensitive to stops because their cross section is smaller.  This search, as well many others that would be sensitive to stops~\cite{Chatrchyan:2011qs, ATLAS:2011ad, Chatrchyan:2011ff,Aad:2011cwa} require at least one lepton in the final state. Other more general searches, including, for example, those for $b$'s, jets, and $\met$~\cite{CMS-PAS-SUS-11-006,ATLAS-CONF-2012-003}, and those for jets and $\met$~\cite{Chatrchyan:2011zy,Aad:2011qa} will also be sensitive to direct stop production, but they will certainly not be optimized.  

Below, we present a method to find purely hadronic stop decays with large $\met$ using fat jets and modern top-tagging algorithms.  Not requiring a lepton makes the search sensitive to the final state with the largest branching ratio.  By exploiting the properties of top decays, this search will be more sensitive to stops than naive searches with jets. Theoretical work looking at this final state both without~\cite{Meade:2006dw,Matsumoto:2006ws,Nojiri:2008ir,Plehn:2010st} and with leptons~\cite{Han:2008gy, Freitas:2006vy,Plehn:2011tf} studied the 14 TeV LHC with a much larger data set.  There is also an analysis focusing on a lower energy LHC which uses leptonic channels~\cite{Bai:2012gs}, and here we present a strategy optimized for near-term discovery with fully hadronic decays.

While completing this work, a similar analysis~\cite{Plehn:2012ab} was published which uses many of the same methods. That analysis uses quantitatively different cuts than we do here, and it focuses mainly on 2012 LHC data, while we will look at both 2011 and 2012 data.

\section{Relevant Processes}

The signal we consider is 
\be
p p \rightarrow \tilde{t} \tilde{t}^* \rightarrow (t \chi^0) \; (\bar{t} \chi^0) \rightarrow (b j j \chi^0) \; (\bar{b} j j \chi^0),
\ee
where $\chi^0$ is a neutral long lived particle and $j$ is a light flavor jet.  We consider only one stop, with the second one potentially much heavier.  We simulate the signal using MadGraph 5~\cite{Alwall:2011uj} and hadronize with Pythia 6~\cite{Sjostrand:2006za}. We use the tree level cross section from Madgraph for the signal and background.  We do not use k-factors to estimate the next-to-leading order effects -- the backgrounds these searches are most sensitive to are in special corners of phase space, and thus k-factors may represent a poor approximation.

There are two classes of backgrounds; those that include top quarks and those that do not. Backgrounds which do not contain tops can be much more easily eliminated with loose ``top-tags,'' and we will discuss them below. The dominant backgrounds, are those with on-shell tops.  The largest of these is $t\bar t + {\rm jets}$, where one of the tops decays leptonically producing a hard neutrino. A lepton veto can eliminate most of the events where the $W$ decays to an electron or muon, but a $\tau$, especially one that decays hadronically, cannot be eliminated in this way.  We simulated background samples of $t\bar t$ and $t\bar t$ + 1 and 2 jets again using MadGraph 5 and Pythia 6.  The samples were matched using the MLM procedure~\cite{Caravaglios:1998yr} with a matching scale of 30 GeV. 

Single top $+\, W\,+$ jets is also a significant background, where one of the two $W$'s decays leptonically.  Extra radiation in these events can make it look like a $t\bar t\;+\met$ event, and we simulate $t+W$ + 1 and 2 jets using the same techniques as the $t \bar t$ +  jets sample.  

The largest background without tops is $Z$ + jets, where the $Z$ decays invisibly.  Since our signal contains fully hadronic tops, the jets produced in association with the $Z$ can in principle look like tops.  As we will see in the next section, this does not happen very often, namely top-tags efficiently suppress this background.  Furthermore, typical $Z$'s are produced with a transverse momentum $\sim M_Z$, so our large $\met$ cut further reduces this background. An even smaller background is $W$ + jets where the $W$ decays leptonically but the charged lepton is either lost or a $\tau$. This is small for the same reasons as $Z$ + jets, and further reduced because in the $W$ decay, the neutrino does not carry all of the $W$ momentum, so an even smaller fraction of the events pass the $\met$ cut. Because we will require a $b$-tag, we separately simulate $V +\, b\bar b \;+$ 1 and 2 jets, and $V+$ 3 and 4 jets, where $V=W,Z$.  

Additional backgrounds which can produce this signal topology include $t\bar t Z$, and diboson production, but the cross sections for these processes are very small and they need not be considered with the amount of luminosity we study here.  

Finally, we note that pure QCD multijet production can produce $\met$ from various sources including leptonic heavy flavor decays and detector effects.  The contribution of these events can be estimated in data (see for example \cite{Aad:2011qa, Chatrchyan:2011bj}) and are small for large $\met$ and events with at least one $b$-tag.  Furthermore, we implement a cut on $\met/\sqrt{\sum E_T}$ as per ATLAS~\cite{ATLAS-CONF-2012-037} and a cut on $\Delta\phi_{\rm min}$  as used, for example, in this CMS study~\cite{CMS-PAS-SUS-11-004}. These cuts have negligible effects on the signal and are described in Section~\ref{sec:signal_extraction}. 

The cross sections for all the signals and backgrounds at both 7 and 8 TeV energies are shown in Table~\ref{tab:xsec}. 

\begin{table}[b]
\caption{Tree level cross sections of signal and relevant background processes. The number in parentheses for the signal rows is the stop mass. The cross section does not depend on the neutralino mass. In $t \bar t$ and single top we require exactly one $W$ to decay leptonically.  Here $V=Z,W$.  }
\begin{tabular}{|c|c|c|c|}
\hline
Process & Generator cuts & $\sigma$ (fb) & $\sigma$ (fb) \\ 
 & and parameters & 7 TeV & 8 TeV \\ \hline 
$\tilde{t} \tilde{t }^*$ ($340$ GeV)& \multirow{3}{*}{$\tilde{t} \tilde{t }^*\rightarrow b \bar{b} +4j + 2\chi$}& 254 &$1.04 \times 10^3$  \\
$\tilde{t} \tilde{t }^*$ ($440$ GeV)& & 48.8 & 205  \\
$\tilde{t} \tilde{t }^*$ ($540$ GeV) & & 11.8 & 51.1   \\ \hline
$t \bar t \,+ $ jets   & $W_{ \rightarrow \ell \nu}, \; p_{T_\nu}>80 \gev $ & $16.3 \times 10^3$ & $26.7 \times 10^{3}$ \\ 
sing.~top  + jets& $p_{T_\nu} > 100 $ GeV & $4.65 \times 10^3$  & $8.27 \times 10^3$ \\ \hline
$V +\, b\bar b\,+$ jets & $Z \rightarrow \nu\bar\nu$, $W \rightarrow \ell \nu $ & $ 1.08\times 10^3 $ & $1.53 \times 10^3$ \\
$V +$ jets  & $\sum {\bf p}_{T_\nu}  > 80$ GeV & $66.6 \times 10^3$ & $96.3 \times 10^3 $  \\   \hline
\end{tabular}
\label{tab:xsec}
\end{table}

\section{Our method of signal extraction}
\label{sec:signal_extraction}

Our event selection is as follows.  We cluster all hadronic activity in the event with $|\eta| < 2.5$ into ``fat'' jets using the FastJet~\cite{Cacciari:2011ma} implementation of the Cambridge-Aachen algorithm~\cite{Dokshitzer:1997in,Wobisch:1998wt} with R=1.2.  Then we take the same event and cluster ``skinny'' jets to mimic the usual experimental algorithms using the anti-$k_t$ algorithm~\cite{Cacciari:2008gp} with R=0.5. We then make the following preselection cuts on our simulated samples which will have limited effect on the signal.\footnote{These pre-cuts will eliminate some signal events, but the vast majority of those events would not have passed our subsequent selection cuts.}

\begin{itemize}

\item We veto all events with an isolated lepton. The isolation criteria is that the lepton's energy comprise at least 80$\%$ of all energy in a cone of $R=0.4$ centered on the lepton. We require $|\eta|<2.5$ and a minimum $p_T$ of 4 and 8 GeV for muons and electrons, respectively.

\item We veto all events that contain a hadronic ``$\tau$''.  We parameterize experimental $\tau$-tagging efficiency by adopting a $50\%$ tagging efficiency for hadronic tau decays in which the hadronic decay products have $p^{had}_T > 20$ GeV and a fake rate of $2\%$ for other jets ~\cite{CMS-PAS-TAU-11-001, ATLAS-CONF-2011-152}. Our tau identification uses skinny jets in order to match the experimental measurement. We assume that hadronic tau decays with less than 20 GeV of visible $p_T$ cannot be tagged.

\item We define $\met = -\sum_i \vec{p_{T_i}}$ where $i$ runs over all visible particles and $H_T=\sum_j p_{T_j}$ where $j$ runs over all skinny jets with $p_T>25$ GeV.  We then require all events to have  $\;\met/\sqrt{H_T} > 5 \sqrt{\rm GeV}$. 

\item We require that the $\met$ vector be separated by more than $\Delta\phi$ of 0.4 from each of the three hardest leading skinny jets.  Namely, $\min[\Delta\phi (\met, j_{i})] >0.4$.  This cut quantifies the notion that $\met$ arising in QCD events tends to lie along a jet direction.

\item We require $H_T>275 \;\gev$. This has practically no effect on the signal since we expect two hadronic top decays.

\end{itemize}
The first two cuts reduce leptonic contributions from $W$ decays. The signal, of course, also can have leptonic $W$ decays and this has been used to look for stops by the experiments~\cite{Aad:2011wc,Chatrchyan:2011qs, ATLAS:2011ad, Chatrchyan:2011ff,Aad:2011cwa}.  Here we are doing the orthogonal search of looking for all hadronic final states so we veto leptons.  The last two cuts are designed to eliminate contributions from pure QCD, and these are cuts that the experiments use for this purpose.  In particular, the resolution on $\met$ in multi-jet events has been shown to be proportional to $\sqrt{H_T}$~\cite{ATLAS-CONF-2012-037}, so requiring large  $\met/\sqrt{H_T}$ reduces QCD events where the $\met$ is mis-measured.  

The intuition for the fourth cut is that large $\met$ in multi-jet events often comes from large mis-measurement of the energy, but not angle, of a single jet. In that case, the $\met$ should point in nearly the same $\phi$ direction as that jet, so requiring $\met$ to be well separated from all the leading jets reduces the QCD background. This variable is used in many hadronic searches, see for example ~\cite{CMS-PAS-SUS-11-004}.

After the preselection, we make our main cuts to distinguish signal from background.

\begin{itemize}

\item We require $\met > 175$ GeV.

\item Among the two highest $p_T$ fat jets, we require one to pass a HEPTopTagger~\cite{Plehn:2010st}. The definition of the algorithm is given in Appendix~\ref{app:top-tagger}.

\item We require the fat jet that is not top-tagged to be $b$-tagged.  We assume a 70$\%$ $b$-tagging efficiency and a 1$\%$ fake rate~\cite{CMS-PAS-BTV-11-001, ATLAS-CONF-2011-102}. As discussed below, we filter the non top-tagged fat jet, and it is the filtered subjets that we use for $b$-tagging. This procedure facilities a more direct comparison with experimental procedures; it would be interesting to have direct measurements of $b$-tagging in fat jets without using subjets.

\end{itemize}
The $\met$ is the hallmark of our signal and is used to distinguish it from events with just jets.  This cut ensures that at least one hard neutrino is present in the background sample (up to drastic mis-measurement, which is accounted for in the preselection cuts). The HEPTopTag helps eliminate $V+\,b\bar{b}\, +$ jets, while the $b$-tag reduces the $V+$ jets background.  We find that requiring the $b$-tag to be in the jet that is not top-tagged improves signal to background relative to allowing either fat jet to be $b$-tagged.  This is because in the dominant background, the fat jet which is not top-tagged is often made up of external radiation rather than real top decay products, so this method of $b$-tagging does a better job of suppressing this background. An interesting alternative to $b$-tagging may be to use a $W$-tag on a non-HEP tagged jet, but we did not explore this possibility it in detail.

\begin{figure}
\includegraphics[width=0.4\textwidth]{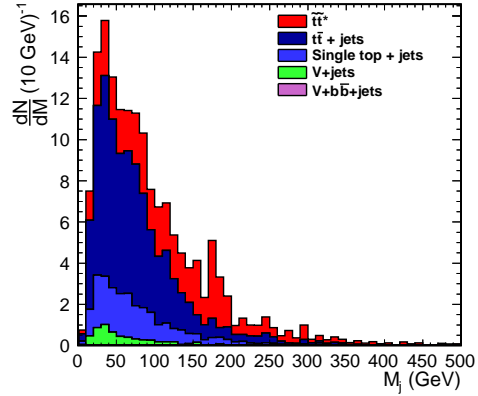}
\caption{Distribution of the non-HEP-tagged filtered jet mass for $(m_{\tilde{t}},m_{\chi})=(340 \; \gev, 0 \; \gev)$ at $\sqrt{s} = 7$ TeV and ${\cal L}=5 \; {\rm fb}^{-1}$. This includes the preselection cuts as well as the main cuts, but not the transverse mass cuts.  }
\label{fig:main_cuts1}
\end{figure}

\begin{figure}
\includegraphics[width=0.4\textwidth]{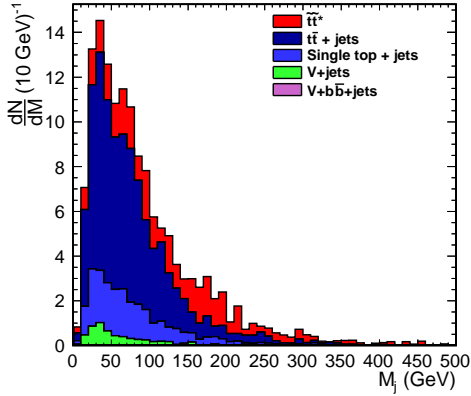}
\caption{Same as Fig.~\ref{fig:main_cuts1} except with $(m_{\tilde{t}},m_{\chi})=(340 \; \gev, 100 \; \gev)$.}
\label{fig:main_cuts2}
\end{figure}

After making these cuts, the dominant background remaining comes from tops, specifically
\be
t\bar{t} + {\rm jets}  \rightarrow (b j j) \; (b \,\tau_h\, \nu_\tau) + {\rm jets},
\label{eq:bg}
\ee
where $\tau_h$ is a hadronically decaying $\tau$. In order to pass the large $\met$ cut, the $\nu_\tau$ typically carries most of the momentum of its mother $W$, although the neutrino produced in the $\tau_h$ decay also has some of the momentum. After making these cuts, we can look at the filtered~\cite{Butterworth:2008iy} invariant mass of the fat jet which is not top-tagged. We have chosen the filtering clustering radius and number of subjets to be $R_{\rm filt}=0.3$ and $n_{\rm filt}=3$.\footnote{We tested pruning~\cite{Ellis:2009su} and trimming~\cite{Krohn:2009th} as well, and found filtering to give a somewhat more peaked distribution about the top mass.} This variable is plotted in Fig.~\ref{fig:main_cuts1}.  In that figure we see that there are some background events that pass the above cuts, but the mass distribution is smoothly falling.  If there are relatively light stops and very light neutralinos, there is a distinct signal peak above the background around the top mass.  Figure~\ref{fig:main_cuts2} shows that increasing the neutralino mass with fixed stop mass reduces the size of the signal peak because there is both less $\met$, and the tops are less boosted on average, so fewer signal events pass our cuts.  Even in Fig.~\ref{fig:main_cuts2}, however, there is still modest evidence for signal events above the background. 

In order to further reduce background, we use two kinematic variables.  First, the usual transverse mass, $m_T$, which is a function of two objects, one of which could be the $\met$ of the event:
\begin{equation}
m^2_T({\bf p}^{\alpha}_T, m_{\alpha}, {\bf q}^{\beta}_T , m_{\beta}) = m^2_{\alpha}+m^2_{\beta} +
         2(E^{\alpha}_T E^{\beta}_T-{\bf p}^{\alpha}_T \cdot {\bf q}^{\beta}_T)
\end{equation}
with
\begin{equation}
E^\alpha_T = \sqrt{m_\alpha^2 + ({\bf p}_{T}^{\alpha})^2}.
\end{equation}
We also use the $m_{T_{2}}$~\cite{Lester:1999tx,Barr:2003rg} variable which is a function of two visible objects as well as the $\met$ of the event and a candidate mass for the $\met$, $m_\chi$:
\begin{align}
&m^2_{T_{2}}({\bf  p}^{\alpha}_T,m_{\alpha}, {\bf p}^{\beta}_T, m_{\beta}, \not {\bf p}_T, m_\chi) = 
\min_{\not{\bf q}^{(1)}_T + \not {\bf q}^{(2)}_T =\not {\bf p}_T}  \nonumber \\
& \left[ \max\{m^2_T({\bf p}^{\alpha}_T, m_{\alpha}, \not {\bf q}^{(1)}_T , m_{\chi}),\,
m^2_T({\bf p}^{\beta}_T,  m_{\beta},  \not {\bf q}^{(2)}_T , m_{\chi}) \}\right].
\end{align}
We have found that the optimal cuts to reduce the dominant background (\ref{eq:bg}) are:
\begin{itemize}

\item Construct $m_{T_{2}}$ with the two leading fat jets and the missing energy of the event. Take the masses of the fat jets to be the measured masses and the missing energy to be massless, $m_\chi=0$.  We require $m_{T_{2}} > 200$ GeV. This computation is done using the algorithm implemented by the  Oxbridge MT2 library~\cite{oxbridge}. 

\item We require that $m_T$ between the $\met$ and each of the leading fat jets be greater than 200 GeV.  Namely, $\min[m_T (\met, j_{1,2})] >200$ GeV. We take the $\met$ to be massless, and the mass for the fat jets to be the measured masses. 

\end{itemize}
The $m_{T_{2}}$ variable is designed for the events in which two identical massive particles decay into two identical invisible particles, as is the case for our signal process.  In events where this is not the case, like the majority of the backgrounds, the $m_{T_{2}}$ value for the event tends be smaller than signal events.  Therefore, this cut on $m_{T_{2}}$ increases the signal to background ratio, and this fact has been exploited in many similar contexts, for example~\cite{Meade:2006dw,Plehn:2010st}.

A second observation is that in the dominant $t\bar t$ background (\ref{eq:bg}), the invariant mass of the hard neutrino with one of the fat jets approaches (but shouldn't be larger than) $m_{\rm top}$ because the neutrino is carrying most the momentum of the $W$. Of course, we can only measure the transverse momentum of the neutrino, but the transverse mass of the neutrino with one fat jet has a cutoff at $\sqrt{m_{\rm top}^2 - m_W^2} \sim 155$ GeV.  This cutoff is observed at parton level, but it is smeared out by hadronization and combinatoric background.  We still find that the cut we implement above significantly improves signal to background ratio. 

\begin{figure}
\includegraphics[width=0.4\textwidth]{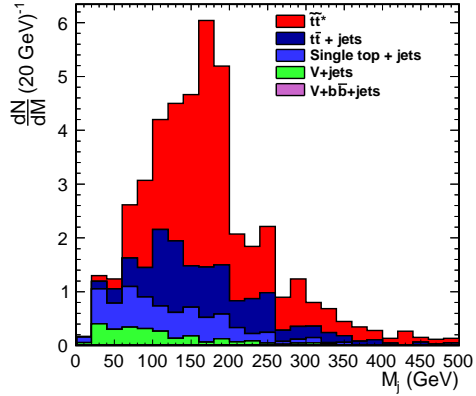}
\caption{Same variable as Fig.~\ref{fig:main_cuts1}, but with the additional kinematic cuts described in the text also imposed, and the binning changed.}
\label{fig:all_cuts1}
\end{figure}

\begin{figure}
\includegraphics[width=0.4\textwidth]{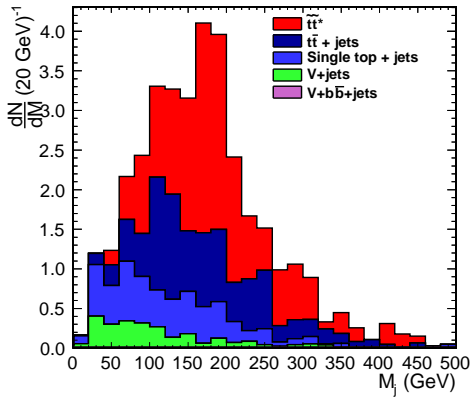}
\caption{Same as Fig.~\ref{fig:all_cuts1} but with  $(m_{\tilde{t}},m_{\chi})=(340 \; \gev, 100 \; \gev)$.}
\label{fig:all_cuts2}
\end{figure}

The same distribution as Fig.~\ref{fig:main_cuts1} is shown in Fig.~\ref{fig:all_cuts1} after implementing the above two kinematic cuts. Comparing Figs.~\ref{fig:main_cuts1} and~\ref{fig:all_cuts1}, we see that the kinematic cuts reduce the number of both signal and background events, but they make the signal jump out above the background. Furthermore, the effect of these cuts on the signal point shown in Fig.~\ref{fig:main_cuts2} is shown in Fig.~\ref{fig:all_cuts2}.  While Fig.~\ref{fig:main_cuts2} shows only modest evidence of a signal, Fig.~\ref{fig:all_cuts2} demonstrates that the kinematic cuts make the excess much more significant.  The detailed results of all the cuts on the backgrounds and some signal points are shown in Table~\ref{tab:passcuts}.

\begin{table*}
\caption{Number of events passing a given cut with 5 ${\rm fb}^{-1}$ luminosity at 7 TeV. The LSP is assumed to be massless.}
\begin{tabular}{|c|c|c|c|c||c|c| }
\hline
Process & Pre-cut  & $\met > 175 $ GeV  &  1 top-tag & $b$-tag & $m_{T_{2}}>200 \; \gev$ & $m_T >200 \; \gev$   \\ \hline 
$\tilde{t} \tilde{t }^*$ ($340$ GeV) &$688$ & 327 & 109 & 50 & 32 & 26 \\
$\tilde{t} \tilde{t }^*$ ($440$ GeV) & 150 & 112 & 42 & 20 & 16 & 14\\
$\tilde{t} \tilde{t }^*$ ($540$ GeV)& 39 & 33 & 13 & 7 & 6 & 6\\ \hline
$t \bar{t} \,+ $ jets  &  $12.5 \times 10^3$&  872&  248 &110 &28  & 18\\ 
Single top + jets & $1.56 \times 10^3$ & 611 & 145 &23  & 8 & 6 \\ 
$V + b \bar{b}\, + $ jets  & 906 &169  & $<1$ & $<1$ &$\ll 1$  &$\ll 1$ \\  
$V \,+ $ jets  & $9.01 \times 10^3$ & $2.34 \times 10^3$ & 166 & 6 & 3 & 2\\ \hline
Total Background & $23.9\times 10^3$ & $3.98 \times 10^3$  &  559 &140  &39 & 27 \\ \hline
\end{tabular}
\label{tab:passcuts}
\end{table*}

\section{Conclusion: Estimated Reach}

\begin{figure}
\includegraphics[width=0.4\textwidth]{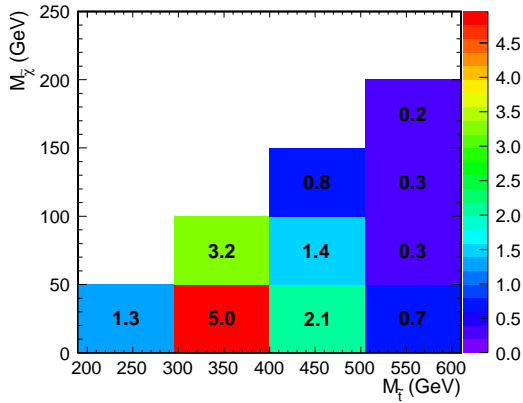}
\caption{Significance of excess computed by counting events in $m_{j_2} \in [150 \; \gev, 230 \; \gev]$ and assuming a Poisson Distribution. This is for $\sqrt{s} = 7$ TeV and $\mathcal L = 5 \, {\rm fb}^{-1}$.}
\label{fig:7tevreach}
\end{figure}

We estimate the reach in the $(m_{\tilde{t}},m_{\chi})$ plane by looking at the invariant mass of the fat jet which is not top-tagged in the events where all cuts have been applied.   This is what is shown in Figs.~\ref{fig:all_cuts1} and~\ref{fig:all_cuts2}. In this sample, we take the events which lie in a top-like mass window of $ [150 \; \gev, 230 \; \gev]$ and do a simple counting experiment of the number of total events versus the number expected background events. We then compute a significance assuming a Poisson distribution.  The results for the 2011 data set of $\sqrt{s} = 7$ TeV and ${\cal L} = 5\; {\rm fb}^{-1}$ are shown in Fig.~\ref{fig:7tevreach}, and we see that if stops exist in this scenario with a mass around 340 GeV along with a very light neutralino, then they can be discovered with current data.  The same data set can also exclude stops up to about 440 GeV. 

\begin{figure}
\includegraphics[width=0.4\textwidth]{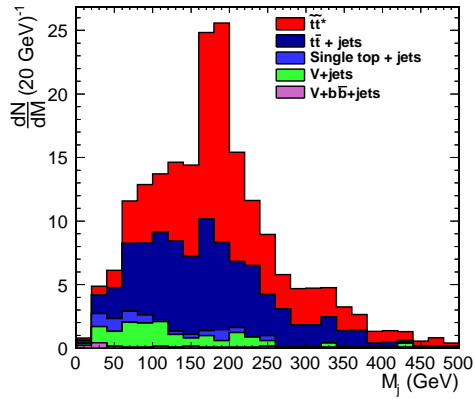}
\caption{Same distribution as Figs.~\ref{fig:all_cuts1} and~\ref{fig:all_cuts2}, but now with $\sqrt{s} = 8$ TeV, ${\cal L} = 20\; {\rm fb}^{-1}$, and $(m_{\tilde{t}},m_{\chi})=(440 \; \gev, 100 \; \gev)$}
\label{fig:mass8tev}
\end{figure}

\begin{figure}
\includegraphics[width=0.4\textwidth]{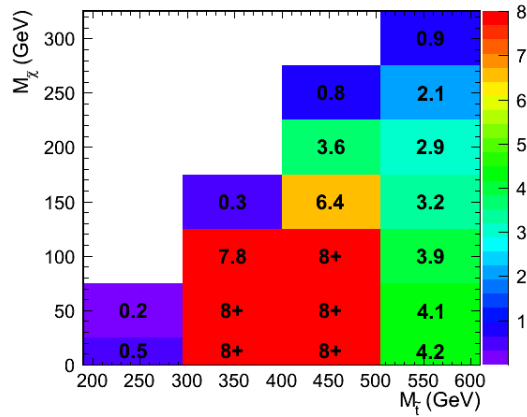}
\caption{Same as Fig.~\ref{fig:7tevreach} but with $\sqrt{s} = 8$ TeV and $\mathcal L = 20 \, {\rm fb}^{-1}$. Boxes with significance $\geq 8$ all have the same color. }
\label{fig:8tevreach}
\end{figure}

Looking forward, we can repeat the analysis for a hypothetical 2012 data set with $\sqrt{s} = 8$ TeV and ${\cal L} = 20\; {\rm fb}^{-1}$. In Fig.~\ref{fig:mass8tev}, we see that even for larger stop and neutralino masses, this data set is enough to see a dramatic signal.  We estimate the reach using the same procedure as for 7 TeV without combining 7 and 8 TeV data sets, and our results are shown in Fig.~\ref{fig:8tevreach}. As expected, the reach improves significantly, and much of the parameter region up to 600 GeV can be covered.

Our work does not take into account systematic errors including uncertainties in the total cross section for the background. This can be ameliorated, however, by doing a more sophisticated shape analysis of the distributions in Figs.~\ref{fig:main_cuts1}--\ref{fig:all_cuts2} rather than our naive counting experiment.  In particular, in Figs.~\ref{fig:all_cuts1} and~\ref{fig:all_cuts2}, we see that after all the cuts, the background is relatively smooth while the signal has a district peak.  There will not be many events after all the cuts, but there is also shape information in the data sample without the kinematic cuts, Figs.~\ref{fig:main_cuts1} and~\ref{fig:main_cuts2}, where there will be many more events. 

One possibility to increase the signal efficiency and extend our reach in the $(m_{\tilde{t}},m_{\chi})$ plane is to decrease our relatively high $\met$ cut.  We have required such a high cut in order to reduce QCD background to a level well below signal. It may be possible, however, to model this background using data driven methods and then subtract it off.  In this case, the thresholds on the kinematic cuts could potentially be increased to further reduce the top background and improve the search, especially for heavier neutralino masses.  For example, if the stop is 250 GeV and neutralino is massless, our current cuts pass less than one signal event with 5 fb$^{-1}$ at 7 TeV, while reducing the $\met$ cut to 100 GeV would allow as many 15 as events to pass.

Finally, we note that while we believe this procedure is optimized for finding direct hadronic stop production, existing experimental searches which use a similar philosophy may be sensitive to stop production.  This CMS search~\cite{CMS-PAS-EXO-11-006} with 1 fb$^{-1}$ uses modern top-tagging methods to look for $t\bar t$ resonances.  While the method used in this search is optimized for more boosted tops, it could still be sensitive to this signal.  In particular, looking at their data sample in events with larger $\met$ could reveal evidence for direct stop production.
 
The experimental effort to find stops is of paramount importance to test whether supersymmetry solves the hierarchy problem. While this effort is underway, every attempt should be made to cover all different possible scenarios.  Here we have shown that if the stop decays to an on-shell boosted top along with a neutralino, then hadronic decays of the top can be exploited to explore this regime with current data, and future data will allow for even broader exploration of the possibilities. 

\vskip 0.3 cm

\noindent
\textbf{Note Added:} While completing this work we became aware of two additional theoretical works looking at similar searches at the LHC~\cite{Han:2012:ab, Alves:2012cd}.

\vskip 0.3 cm

\noindent
We thank Eder Izaguirre, Petar Maksimovic, and Salvatore Rappoccio for useful discussions. This work is supported in part by NSF grants PHY-0244990 and PHY-0401513, by DOE grant DE-FG02-03ER4127, and by the Alfred P. Sloan Foundation. DS is supported in part by the NSF under grant PHY-0910467 and gratefully acknowledges support from the Maryland Center for Fundamental Physics.

\begin{appendix}

\section{Description of HEPTopTagger}
\label{app:top-tagger}

When a boosted top decays, its decay products will often be collimated. This fact has been exploited to create several top-taggers~\cite{Thaler:2008ju, Kaplan:2008ie,Almeida:2008yp} that use various handles of hadronic top decays to distinguish them from jets which come from other sources. In stop searches with moderate stop masses such as those considered in this work, the rate for highly boosted tops in the final state is significantly lower than the overall signal rate. Therefore, a top-tagger which works when the tops have only $O(1)$ boost would increase the signal efficiency. We find that the HEPTopTagger works best for modestly boosted tops, and that is why we choose it for this analysis.  

A detailed description of the HEPTopTagger is found in~\cite{Plehn:2010st}. We give a schematic description of the algorithm here for completeness. The algorithm takes as input the fat jets described in Section~\ref{sec:signal_extraction}. It is applied to individual fat jets, and each one either passes or fails. The algorithm is as follows:

\begin{enumerate}

\item Begin by undoing the clustering algorithm to look for subjets. When the mother jet becomes two daughters, if the daughter jet with larger mass has less than 80\% of the mass of the mother jet, keep both jets.  Otherwise, keep just the daughter with larger mass. Continue this process as long as subjets have mass greater than 30 GeV. 

\item Filter sub jets with resolution $R_{\rm filter} = \min(0.3,\Delta R_{jk}/2)$. Keep the five filtered subjets with the highest $p_T$. If there are fewer than five, keep all of them.

\item Select the set of three subjets whose combined mass is closest to $m_{\rm top}$. Require that the combined $p_T$ of the three subjets is more than 200 GeV. 

\item Sort the three selected subjets by $p_T$, $j_1, j_2, j_3$ and calculate pairwise invariant masses, $m_{12}, m_{23}, m_{13}$.  

\item If the masses satisfy 
\begin{equation}
0.2 <\arctan \frac{m_{13}}{m_{12}} < 1.3
\end{equation}
and 
\begin{equation}
R_{\min}< \frac{m_{23}}{m_{123}} < R_{\max},
\end{equation}
where $R_{\min} = 85\%\, m_W /  m_{\rm top}$ and $R_{\max} = 115\%\, m_W /  m_{\rm top}$,
then this is a top candidate.

\item If the masses satisfy
\begin{eqnarray}
R_{\min}^2 \left(1+\left(\frac{m_{13}}{m_{12}}\right)^2 \right)  \nonumber \\
< 1-\left(\frac{m_{23}}{m_{123}} \right)^2     \\
< R_{\max}^2 \left(1+\left(\frac{m_{13}}{m_{12}}\right)^2 \right)  \nonumber
\end{eqnarray}
and
\begin{equation}
\frac{m_{23}}{m_{123}} > 0.35
\end{equation}
then this is also a top candidate.  

\item If the masses satisfy 
\begin{eqnarray}
R_{\min}^2\left(1+\left(\frac{m_{12}}{m_{13}}\right)^2 \right) \nonumber \\
< 1-\left(\frac{m_{23}}{m_{123}} \right)^2  \\
< R_{\max}^2\left(1+\left(\frac{m_{12}}{m_{13}}\right)^2 \right)  	\nonumber
\end{eqnarray}
and
\begin{equation}
\frac{m_{23}}{m_{123}}> 0.35
\end{equation}
this is a top candidate as well.  

\item If the masses do not satisfy any of the three previous conditions, then this jet is not a top candidate. 

\end{enumerate}
A full explanation of why these criteria are chosen and what the efficiency and fake rate of this algorithm are can be found in~\cite{Plehn:2010st}.

\end{appendix}

\end{document}